\documentclass[pre,reprint]{revtex4-1}
\usepackage{amsmath}
\usepackage{graphicx}				
\usepackage{amssymb}

\newcommand*{\pd}[2]{\frac{\partial #1}{\partial #2}}

\newcommand*{\Tr}[1]{\operatorname{Tr}\left[{#1}\right]}
\newcommand*{\Tri}[2]{\operatorname{Tr}_{#1}\left[{#2}\right]}
\newcommand*{\avg}[1]{\left\langle{#1}\right\rangle}
\newcommand*{\tavg}[1]{\langle{#1}\rangle}

\newcommand{\pure}[1]{|#1\rangle\langle #1|}
\newcommand{\bra}[1]{\langle #1|}
\newcommand{\ket}[1]{|#1\rangle}

\begin{document}

\title{ The EPR Paradox Implies A Minimum Achievable Temperature}
\author{ David M. Rogers}
\affiliation{ University of South Florida, Tampa}

\begin{abstract}
  We carefully examine the thermodynamic consequences of the repeated
partial projection model for coupling a quantum system to an
arbitrary series of environments under feedback control.
This paper provides observational definitions of heat and work
that can be realized in current laboratory setups.
In contrast to other definitions, it uses only properties of the environment and
the measurement outcomes, avoiding references to the `measurement'
of the central system's state in any basis.
These definitions are consistent with the usual laws of thermodynamics at all temperatures,
while never requiring complete projective measurement of the entire system.
It is shown that the back-action of measurement must be
counted as work rather than heat to satisfy the second law.
Comparisons are made to stochastic Schrodinger unravelling
and transition-probability based methods, many of which appear as
particular limits of the present model.
These limits show that our total entropy production is a lower bound on traditional
definitions of heat that trace out the measurement device.
Examining the master equation approximation to the process
at finite measurement rates, we show that most interactions with
the environment make the system unable to reach absolute zero.
We give an explicit formula for the minimum temperature
achievable in repeatedly measured quantum systems.
The phenomenon of minimum temperature offers a novel explanation of recent
experiments aimed at testing fluctuation theorems in the quantum realm
and places a fundamental purity limit on quantum computers.
\end{abstract}

\maketitle

\section{Introduction}

A version of the EPR paradox prevents simultaneously doing work on a quantum system
and knowing how much work has been done.
A system can do work on its environment only if the two have a nonzero
interaction energy.  During interaction, two become entangled,
leading to a superposition of different possible values for the work.
According to quantum mechanics, measuring the work projects
into a state with exactly zero interaction energy.  Therefore the
system-environment interaction is always either zero or unknown.

  One hundred years ago, Einstein presented a first-order rate hypothesis concerning the
rate of energy exchange between a molecular system and a reservoir of photons.\cite{aeins16}
Under this hypothesis, the transition between states with known molecular energy levels
by emission and absorption of discrete photons can be shown to bring about
thermal equilibrium for all parties: the photons, the molecular energy levels,
and the particle velocities.  This semiclassical picture provided a clear,
consistent, and straightforward picture for the time-evolution of coupled
quantum systems.  Nevertheless, the argument
must have appeared unsatisfactory at the time because it only provided
a statistical, rather than an exact, mechanical description of the dynamics.

  Many years later, Einstein, Podolsky, and Rosen published the famous EPR paradox.\cite{epr,epr2}
The paradox states that, before any measurement is made, neither position nor velocity
exist as real physical quantities for a pair of entangled particles.  Either of the two choices can
be `made real' only by performing a measurement.
The consequence for energy exchange processes follows directly.
For a particle entangled with a field, neither a definite
(molecular energy level / photon number) pair nor a definite (Stark state / field phase)
pair exist before any measurement is made.

  Recent works on quantum fluctuation theorems confront this difficulty in a variety
of ways.  One of the most prominent is the stochastic Schr\"{o}dinger equation
that replaces a dissipative quantum master equation with an ensemble of trajectories
containing periodic jumps due to measurement.\cite{jhoro12}  In that setup, the jump
process represents dissipation, so heat is defined as any energy
change in the system due to the jumps.  Other changes in energy, caused by varying the
Hamiltonian in time, are counted as work.  Fluctuation theorems for this process
are based on the detailed balance condition for jumps due to the reservoir,
avoiding most issues with defining a work measurement.

  The work of Venkatesh\cite{bvenk15} shows that regular, projective measurement
of work-like quantities based on the system alone (such as
time-derivative of the Hamiltonian expectation)
generally leads to ``qualitatively different statistics
from the [two energy measurement] definition of work
and generally fail to satisfy the fluctuation relations of
Crooks and Jarzynski.''

  Another major approach is to model the environment's action as a series of
generic quantum maps.  A physical interpretation as a two-measurement
process accomplishing feedback control was given by Funo.\cite{kfuno13}
There, an initial partial projection provides classical information that is used
to choose a Hamiltonian to evolve the system for a final measurement.  That work
showed that the transition probabilities in the process
obey an integral fluctuation theorem.  Although the interpretation relied on a final
measurement of the system's energy, it provided one of the first examples for
the entropic consequences of measurement back-action.\cite{sdeff16}

  Recent work on the statistics of the transition process for general quantum maps
showed that the canonical fluctuation theorems hold if
the maps can be decomposed into transitions between
stationary states of the dynamics.\cite{gmanz15}
This agrees with other works
showing the importance of stationary states in
computing entropy changes from quantum master equations.\cite{rkosl13}
The back-action due to measurement is not present in this case.

  In contrast, the present work starts from a physically motivated process
and shows that work and heat can be defined without
recourse to stationary states of the central system.
By doing so, it arrives at a clear picture of the back-action, and
a minimum temperature argument.  It also builds a quantum parallel
to the measurement-based definition of work and heat
for classical nonequilibrium systems laid out in Ref.~\cite{droge12}.
There, the transition probability ratio is shown to be equivalent to
a physical separation of random and deterministic forces.
Although no fluctuation theorem can be shown in general,
in the van Hove limit, the interaction
commutes with the stationary state,\cite{bvenk15}
and a fluctuation theorem such as the one in Ref.~\cite{gmanz15}
applies.

  Our model uses a combination of system and reservoir with joint Hamiltonian,
\begin{equation}
\hat H = \hat H_A + \hat H_B + \gamma \hat H_{AB}
. \label{e:en}
\end{equation}
The coupling Hamiltonian should not be able to simply
shift an energy level of either system,
which requires $\Tri{A}{f(\hat H_A) \hat H_{AB}} = 0$ and $\Tri{B}{f(\hat H_B) \hat H_{AB}} = 0$,
for arbitrary scalar functions, $f$.  A simple generalization discussed later
is to waive the first constraint, but this is not investigated here.

  There have been many definitions proposed for heat and work in quantum systems.
These fall roughly into three categories: the near-equilibrium limit, experimental work-based
definitions, and mathematical definitions based on information theory.

  The near-equilibrium limit is one of the earliest models, and is based
on the weak-coupling limit of a system interacting with a quantum energy reservoir
at a set temperature over long time intervals.
  That model is probably the only general one derivable from first principles
where it can be proven that every system will eventually relax to
a canonical equilibrium distribution with the same temperature as the reservoir.\cite{hspoh78}
The essential step is taking the van Hove limit,
where the system-reservoir interaction energy scale, $\gamma$, goes to zero (weak coupling)
with constant probability for energy-conserving transitions (which scale as
$\gamma^2/(\hbar^2 \lambda)$).
In this limit, the only allowed transitions are those
that conserve the uncoupled energy, $\hat H_A+\hat H_B$.
The dynamics then becomes a process obeying detailed-balance for hopping between
energy levels of the system's Hamiltonian, $\hat H_A$.  States with energy superpositions
can mix, but eventually decay to zero probability as long as the environment can couple to every
system energy level.

  Adding an effective time-dependent Hamiltonian, $\hat H^\text{eff}_A(t)$, onto this picture
and assuming very long time-scales 
provides the following definitions of heat and work,\cite{ralic79}
\begin{align}
\dot Q &= \Tr{\hat H^\text{eff}_A(t) \dot \rho} \notag \\
\dot W &= \Tr{\pd{\hat H^\text{eff}_A(t)}{t} \rho} ,\label{e:QW} \\
\intertext{where $\dot F = dF/dt$ denotes the time-derivative of $F$ according to the dynamics,
and $e^{-\beta \hat H^\text{eff}_A(t)}$ must be the stationary state of the time-evolution used.
Note that to match the dynamics of a coupled system, $\hat H^\text{eff}_A(t)$ must be
a predefined function of $t$ satisfying, (see Eq.~\ref{e:dynab})}
\Tr{\hat H^\text{eff}_A(t) \Tri{B}{\rho_{AB}}} &= \Tr{(\hat H_A + \gamma \hat H_{AB}) \rho_{AB}} 
\label{e:ematch}
\end{align}
Work and heat defined by equation~\ref{e:QW} have been used
extensively to study quantum heat engines.%
\cite{ralic79,egeva92,tkieu04,hquan07,mespo10,swook11,ldios11,hli13,rkosl13}
For this definition, it is possible to prove convexity,\cite{hspoh78} and positivity of
$\dot S_\text{tot} = \dot S_A - \beta \dot Q$.\cite{ralic79}
Statistical fluctuations of heat and work have also been investigated.\cite{hquan08,jhoro12,kfuno13,gmanz15}
These first applications have demonstrated some of the novel properties of
quantum systems, but encounter conceptual difficulties when applied to
dynamics that does not follow the instantaneous eigenstates of $H^\text{eff}_A(t)$.\cite{rkosl13,bvenk15,sdeff16}

  The paradox described in this work shows why moving away from eigenstates is so difficult.
The small-coupling, slow-process limit under which Eq.~\ref{e:QW} applies also amounts
to an assumption that the system-environment pair is continually being projected
into states with known $\hat H_A + \gamma \hat H_{AB}$.
It is not suitable for use in deriving modern fluctuation theorems because its
validity relies on the this limit.

  Entropy can also be defined thermodynamically by analyzing physical processes
taking an initial state to a final state.
One of the simplest results using the thermodynamic approach is that even quantum
processes obey a fluctuation theorem for exchanges of (heat) energy between
system and environment when each transition conserves energy and there is no
external driving force.\cite{cjarz04}
On averaging, this agrees with the common experimental definition of heat production
as the free energy change of two reservoirs set up to dissipate
energy by a quantum contact that allows monitoring the energy exchange process.\cite{yutsu10,yutsu12,jkosk13,jpeko15}
Semiclassical trajectories have also been investigated as a means
to show that postulated expressions for quantum work
go over to the classical definition in the high-temperature or small-$\hbar$ limit.\cite{cjarz15}

  Other works in this category consider a process where the system's
energy is measured at the start and end of a time-dependent driving process.
It is then easy to show that the statistics of the energy change give
a quantum version of the Jarzynski equality for the free energy difference.\cite{htasa00,ptalk07a}
More general results are difficult owing to the fact that, for coupled systems,
quantum transitions that do not conserve energy are possible,
giving rise to the paradox motivating this work.


  There have also been many mathematically-based definitions of entropy production
for open quantum systems.
The primary goal of a mathematical definition is
to quantify the information contained in a quantum state.\cite{vvedr02}
It is well-known that preparation of a more ordered system state
from a less ordered one requires heat release proportional to the information
entropy difference.\cite{elutz15,jparr15}
From this perspective, information is more fundamental than measured heats,
because it represents a lower bound on any physical process that could
accomplish this transformation.
A maximum work could be found from such a definition
using energy conservation.
However, the disadvantage of a mathematical
definition is that it can not be used to construct a physical transformation process
obeying these bounds.

  Most of the bounds on mathematical entropy production are proven with the help of the Klein inequality
stating that relative entropy between two density matrices must be positive.\cite{mrusk90}
There are, in addition, many connections with communication and measure theory
that provide approximations to the relative entropy.\cite{vvedr02,tsaga13}

  One particular class of mathematical definitions that has received special attention
is the relative entropy,
\begin{align}
S(\rho | \rho^\text{inst}) &= \Tr{\rho \log \rho - \rho \log \rho^\text{inst}} \notag \\
&= \beta (F(t) - F^\text{(eq)}) \label{e:Srel} \\
\intertext{ between an arbitrary density matrix and an `instantaneous
equilibrium' state,}
\rho^\text{inst} &= \exp{\left[-\beta \hat H^\text{eff}(t)\right]}/Z^\text{eff}(\beta,t)
.\label{e:pos}
\end{align}
This definition is closely related to the physical process of measuring the system's
energy at the start and end of a process.
Several notable results have been proven in those works, including work relations
and integrated fluctuation theorems\cite{htasa00,mcamp09,mcamp09a,kfuno13,gmanz15}
as well as useful upper and lower bounds.\cite{sdeff10,jhoro12}
The present work is distinguished from these mathematical definitions because
it completely removes the requirement for defining or using an `instantaneous equilibrium'
distribution of the central system or directly measuring the central system at all.

  One of the primary motivations for this work has been to derive a firm
theoretical foundation for analyzing time-sequences of measurements
in hopes of better understanding the role of the environment in decoherence.%
\cite{vbrag80,csun97,ejoos98,wstru00,qturc00a,nherm00,rpolk01,wzure02,mball16,bdanj16}
The present paper provides a new way of understanding the gap between the Lindblad operators
describing the quantum master equation and the physical processes responsible for decoherence.
Rather than unravelling the Lindblad equation, we choose a physical process
and show how a Lindblad equation emerges.
This path shows the importance of the source of environmental
noise in determining the low-temperature steady-state.
The result also provides an alternative continuous time, Monte Carlo method for
wavefunction evolution\cite{jdali92,hcarm93}
without using the dissipation operator associated with the Lindblad master equation.

  Another outcome has been finding a likely explanation for the
anomalous temperature of Utsumi et. al.\cite{yutsu10,yutsu12}
Those works attempted to test the classical fluctuation theorems
for electron transport through a quantum dot, and found that the
effective temperature of 1.37 K (derived from the slope of the transport odds ratio,
$\log p_\text{fwd} / p_\text{rev}$) was much higher than the electron
temperature of 130-300 mK.
Trying to lower the temperature further below that point showed minimal changes
in the slope, indicating a minimum temperature had been reached.

  Sections~\ref{s:process} and~\ref{s:therm} present a repeated measurement process, and show that it allows
for a physical definition of heat and work that occurs between successive
measurements.  Measurements are only performed on the interacting reservoir,
and (because of entanglement) cause instantaneous projection of the central system according to the standard
rules of quantum mechanics.  In this way, it is not required to define a temperature
for the central system.  Because the central system is generally out of equilibrium,
the concept of  equilibrium is applied only to the environmental interactions.

  Section~\ref{s:clausius} proves the Clausius form of the second law for the new definitions,
and section~\ref{s:jcm} immediately applies these to the quantum theory of radiation.
The limits of slow and fast measurement rates are investigated in sections~\ref{s:weak} and~\ref{s:strong}.
The slow rate limit recovers Einstein's picture of
first-order rate processes and complies with
Eq.~\ref{e:QW} when the system-reservoir coupling, $\gamma$, is infinitesimally small.
The fast measurement limit does not exhibit a quantum Zeno paradox,\cite{bvenk15}
but effectively injects white noise into the energy
of the joint system -- consistent with the energy-time uncertainty principle.
At intermediate stages, continuous finite interaction with the reservoir
causes an effective increase in the `temperature' of the system's steady-state.
Although surprising, the measurement rate is unavoidable in the theory
as it is the exact parameter controlling broadening of spectral lines.\cite{epowe93}
I end with a proof in section~\ref{s:mint} that effects from the minimum achievable temperature will
be seen when the reservoir temperature is less than the system's
first excitation energy and the measurement rate is on the order of this excitation energy.

\section{ Repeated Measurement Process}\label{s:process}

\begin{figure}
\includegraphics[width=0.45\textwidth]{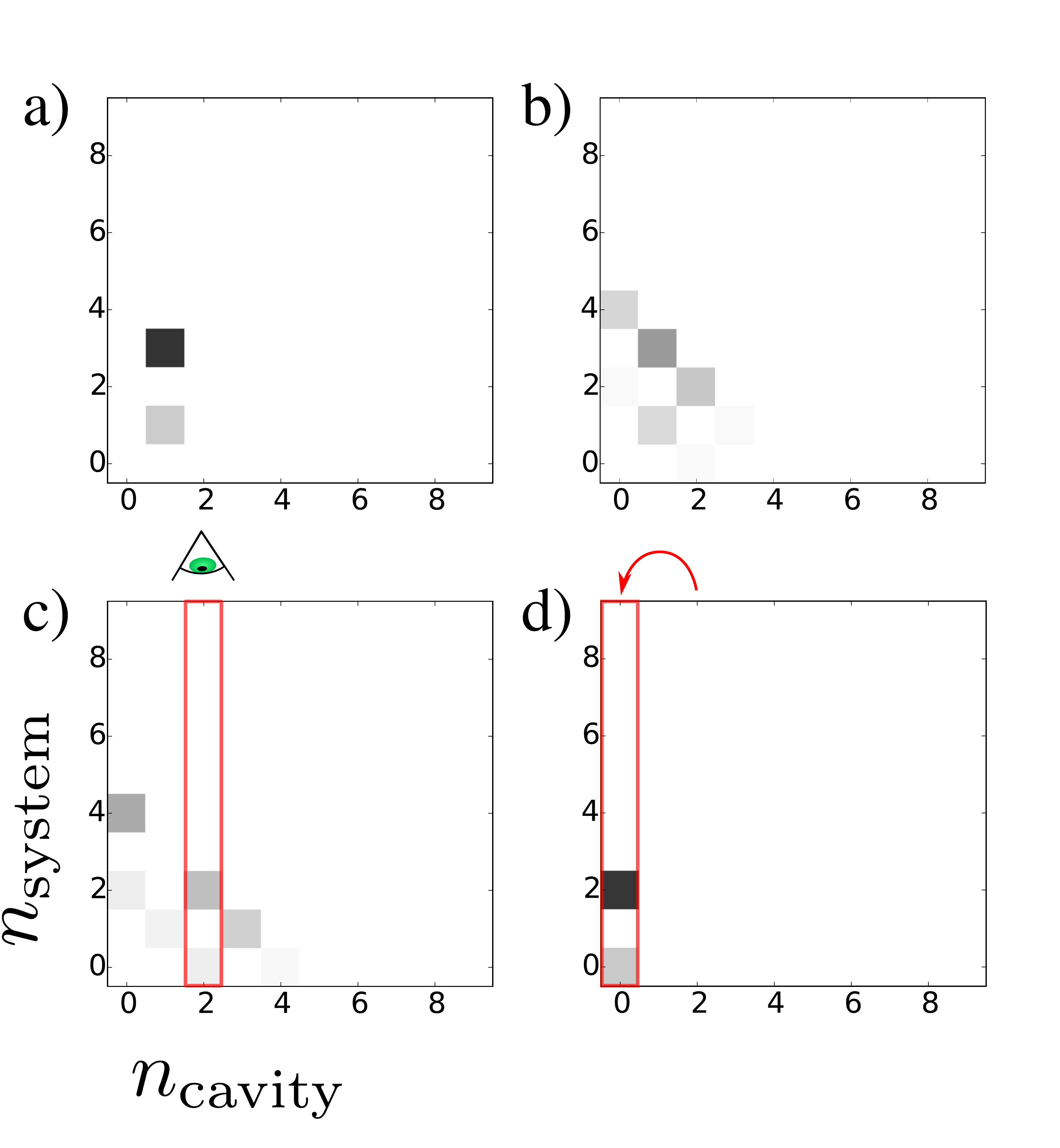}
\caption{Schematic of the repeated measurement process.
(a-c) Exact evolution of the coupled system+reservoir from an uncoupled state
quickly leads to an entangled state.
(c) Measuring the reservoir energy selects a subsample of the system,
removing coherences.
(d) Replacing the reservoir state with a thermal sample results
in heat and work output.  The thermal nature of the environment is
responsible for dissipation.}\label{f:ref}
\end{figure}

  To study the action of continual environmental measurement on
part of a quantum system, I propose the following process (Fig.~\ref{f:ref}):
\begin{enumerate}
  \item Let $\ket{\psi}$ represent a general wavefunction of the central system, and $\ket{n}$ represent
           the state of the measurement device at energy level $\hat H_B \ket{n} = \hbar \omega^B_n \ket{n}$.
  \item The central system is coupled to the measurement device whose state is chosen at random
  from a starting distribution, $\rho_B(0)$, (panel d-a)
  \begin{equation}
  \ket{\psi} \to \ket{\psi}\otimes \ket{n}.
  \end{equation}
   The starting distribution must have a well-defined energy, and so $\rho_B(0)$ should be diagonal
   in the energy basis of system $B$.
  \item The joint system is evolved forward using the coupled Hamiltonian,
  $\hat U(t) = e^{-it \hat H/\hbar}$
  until the next measurement time, chosen
  from a Poisson process with rate $\lambda$ (panel b-c).
  \begin{equation}
  \ket{\psi, n} \to \hat U(t) \ket{\psi, n}
  \end{equation}
  \item The state of the measurement device is `measured' {\em via}
  projection into one of its uncoupled energy eigenstates, $\ket{m}$ (panel c).
  \begin{equation}
  \hat U(t) \ket{\psi, n} \to \frac{\bra{m} \hat U(t) \ket{\psi, n}}{\sqrt{p_m}},
  \end{equation}
  with probability $p_m = |\avg{m|U(t)|\psi, n}|^2$.
\end{enumerate}

  The measurement process itself is described exactly by the `purification'
operator of Spohn and Lebowitz,\cite{hspoh78}
whose effect on the joint density matrix is given by,
\begin{equation}
\hat P \rho_{AB} = (\operatorname{Tr}_B \rho_{AB}) \otimes \rho_B(0)
.\label{e:pure}
\end{equation}
Every time this operation is performed, the memory of the environmental system is
destroyed, and all system-environment superposition is removed.

  For studying thermalization process, it suffices to use a thermal equilibrium
distribution for $\rho_B(0)$,
\begin{equation}
\rho^\text{eq}_B(\beta) = e^{-\beta \hat H_B}/Z_B(\beta)
\label{e:eqB}.
\end{equation}
In many experimental cases, $\rho_B(0)$ represents a specially
prepared input to drive the system toward a desired state.

  The operation of measurement disconnects the two systems, and, more importantly,
makes the energy of the reservoir system correspond to a physical observable.
A complete accounting for heat in quantum mechanics can be made
using only these measurements on ancillary systems, rather than the central, $A$, system.
The thermodynamics based on this accounting allows the central system to retain most of
its quantum character, while at the same time deriving the traditional, operational
relationships between heat and work.

  Although the analysis below is phrased in terms
of density matrices, that view is equivalent to carrying out this process many times with
individual wave-functions.  Specifically, if $\rho_{A}(0) = \sum_j p_j \pure{\psi_j}$,
is composed of any number of pure states,\cite{ejayn57a} the final density matrix at time $t$
is a linear function of $\rho_A$ and hence of each $\pure{\psi_j}$.
Carrying out the process on individual wave-functions thus allows an extra degree
of choice in how to compose $\rho_A(0)$, the use of which does not alter any of the results.

  This process is a repeatable version of the measurement and feedback control
process studied in Ref.~\cite{kfuno13}, and fits into the general quantum map scheme
of Ref.~\cite{gmanz15}.  Nevertheless, our analysis finds different results because the
thermodynamic interpretation of the environment and measuring device
allows the reservoir to preform work in addition to exchanging heat.

\section{ Thermodynamics of Repeated Measurement}\label{s:therm}

\begin{figure*}
\includegraphics[width=0.7\textwidth]{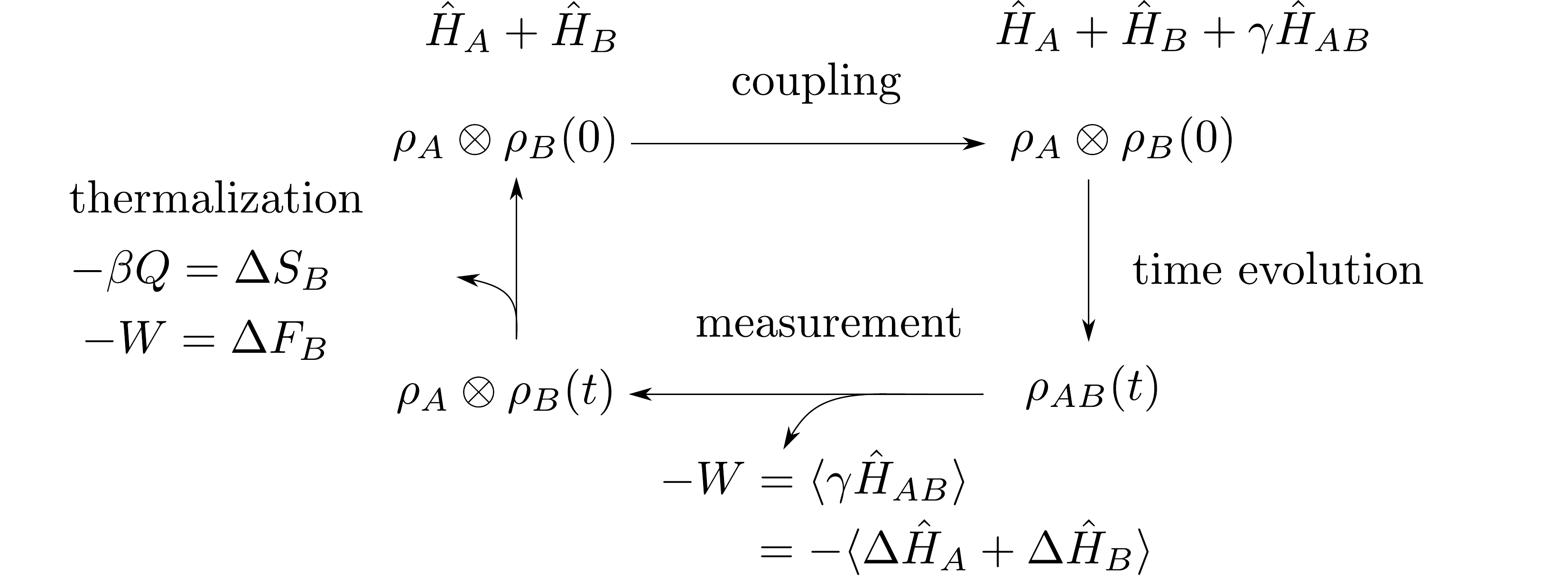}
\caption{Work and Heat of the intermittently measured quantum system.
On the left, the system (A) and reservoir (B) Hamiltonians are uncoupled.
Coupling does not initially change their energy, since diagonal elements of
$\hat H_{AB}$ are zero.  During time-evolution, the total energy is conserved,
leading $\tavg{\hat H_{AB}}$ and $\tavg{\hat H_A + \hat H_B}$ to oscillate.  Measurement
projects back into an uncoupled state, requiring work $-\tavg{\gamma \hat H_{AB}}$.
Finally, thermalization of the reservoir removes accumulated heat, while
exporting all work to the environment.}\label{f:therm}
\end{figure*}

  In order for heat and work to have an unambiguous physical meaning, they must be
represented by the outcome of some measurement.  Fig.~\ref{f:therm} presents
the energies for each operation applied to a system and its reservoir over
the course of each measurement interval in Fig.~\ref{f:ref}.
Initially (in Step 2), the density matrix begins as a tensor product, uncoupled
from the reservoir, which has a known starting distribution, $\rho_B(0)$.
However, for a coupled system
and measurement device, time evolution leads to entanglement.
At the time of the next measurement, the entanglement is projected out,
so it is again permissible to refer to the properties of the $A$ and $B$ systems separately.

  After a measurement, the total energy of the system/reservoir pair
will have changed from $\tavg{\hat H_A + \hat H_B + \gamma \hat H_{AB}}$ to
$\tavg{\hat H_A + \hat H_B}$.
The amount of energy that must be added to `measure' the system/reservoir pair
at any point in time is therefore, $-\gamma \tavg{\hat H_{AB}}$.

  This step is responsible for the measurement `back-action', and the violation
of the FT for general quantum dynamics.
Strictly speaking, this measurement energy does not correspond to an element of physical reality.
Nevertheless, the starting and ending $\hat H_A$, $\hat H_B$ are conserved
quantities under the uncoupled time-evolution, and so the energy of the measurement
step can be objectively defined in an indirect way.

  This instantaneous measurement of the reservoir simulates the physical
situation where an excitation in the reservoir leaks out into the environment.
After this happens, the information it carried is available
to the environment, causing traditional collapse of the system/reservoir pair.

  To complete the cycle, the reservoir degree of freedom must be
replaced with a new sample from its input ensemble.
For the micromaser, this replacement is accomplished spatially by passing separate
atoms ($B$) through a cavity, one at a time.

  On average, the system should output a `hot' $\rho_B(t)$, which the
environment will need to cool back down to $\rho_B(0)$.
Using the methods of ordinary thermodynamics,\cite{ralic79,tkieu04,swook11}
we can calculate the minimum heat and maximum work for
transformation of $\rho_B(t)$ back to $\rho_B(0)$ via an isothermal,
quasistatic process at the set temperature of the reservoir,
\begin{align}
\beta Q &= -\Tr{\rho_B(0) \log \rho_B(0)} + \Tr{\rho_B(t) \log \rho_B(t)} \notag \\
 &= -\Delta S_B \label{e:Q} \\
W_\text{therm} &= \Tr{(\rho_B(0) - \rho_B(t)) \hat H_B} + \Delta S_B/\beta \notag \\
&= - \Delta F_B \label{e:Wtherm} \\
W &= W_\text{therm} + \Delta H_A + \Delta H_B \notag \\
 &= \Delta H_A - Q \label{e:W}
\end{align}
These sign of these quantities are defined as the energy added to the system,
while $\Delta X \equiv \tavg{\hat X}_\text{final} - \tavg{\hat X}_\text{initial}$ represents
the total change in $\hat X$ during evolution from one measurement time to the next.

  In this work, $T$ always refers to the externally set temperature
of the reservoir system.
The temperature of the reservoir, used in defining $\beta = 1/k_B T$ above,
is entirely related to the conditions under which the reservoir states are prepared.
It can be different for each measurement interval.

  Note that when a thermal equilibrium distribution
is used for the reservoir (Eq.~\ref{e:eqB}), the reservoir
dissipates energy from the system.
Since it always begins in a state of minimum free energy,
the reservoir always recovers work from the system as well,
since $-W_\text{therm}$ is always strictly positive by Eq.~\ref{e:pos}.
This makes sense when the central system is relaxing from an initial excited state.
When the central system is at equilibrium, the second law is saved (Sec.~\ref{s:clausius})
by including the work done during the measurement step.

\subsection{ Caution on Using a Time-Dependent Hamiltonian}\label{s:issues}

  The assumption of a time-dependent Hamiltonian for the system leads
to an ambiguity on the scale of the measurement back-action.\cite{kfuno13,bvenk15,sdeff16}
This presentation does not follow the traditional route of assuming a time-dependent
Hamiltonian for the central system.
The assumption of a time-dependent Hamiltonian is awkward to work with in this context
because it side-steps the measurement paradox.
Instead, it assumes the existence of a joint system wherein the dynamics for sub-system $A$
is given exactly by, $\dot \rho_A(t) = -\frac{i}{\hbar}[\hat H^\text{eff}_A(t), \rho_A(t)]$.

  The complete physical system plus environment must have a conserved energy function.
This matches the dynamics,
\begin{equation}
\dot \rho_A(t) = -\frac{i}{\hbar}\Tri{B}{\hat H_A + \hat H_B + \gamma \hat H_{AB}, \rho_{AB}(t)}
\label{e:dynab}
\end{equation}
exactly when Eq.~\ref{e:ematch} holds.

In classical mechanics, such a function can be formally constructed by adding
an ancillary degree of freedom, $y$, that moves linearly with time, $y(t) = t$.
The potential energy function,
\begin{equation}
V(x,y) = V(x) + V^\text{int}(x,y) - \int_0^y \pd{V^\text{int}(x_\text{ref}(t),t)}{t} dt
\end{equation}
is defined using the known trajectory for $x_\text{ref}(t)$
under the desired Hamiltonian, $H(x,t)$, so that
so $y$ experiences no net force.
Alternatively, $y$ can be considered to be infinitely massive.

  When translated to quantum mechanics, neither of these last two methods avoids
the Heisenberg uncertainty principle.\cite{lland3,cjarz15}
An intuitive argument can be based on $\tavg{\Delta p} \tavg{\Delta x} \ge \frac{\hbar}{2}$.
In both cases, the work done by the system on the reservoir is, $\pd{V^\text{int}(x,y)}{dy} dy$,
and contributes directly to the change in momentum of $y$.
The $y$-coordinate was constructed to move linearly in time, and hence
measures the `time' of interaction.
Using these translations from momentum change and position to work / time
provides, $\tavg{\Delta p_y} \tavg{\Delta y} \simeq \tavg{\Delta_t V(x,t)}\tavg{\Delta t}$.

  Although the definitions of heat and work in Eq.~\ref{e:QW} can be
shown to be mathematically consistent with the laws of thermodynamics,
they require infinitesimally slow time-evolution under the
Markov assumption and constant comparison to a steady-state distribution.\cite{ralic79,rkosl13,gmanz15}
The present method is valid under a much less restrictive set of assumptions.
In particular, it allows arbitrary time-evolution,
and only makes use the equilibrium properties of the $B$ system,
not the central, $A$ system.  The present set of definitions is also directly connected
to the experimental measurement process.

  Defining a time-dependent $\hat H$ as is done in other works
groups the central system together with some aspects of the reservoir.
In the present framework, it is easy to allow $\hat H_B$ and
$\hat H_{AB} = \hat H_{AB}^0 + \hat H_A'$ to be different for each measurement interval
(encompassing even non-Markovian dynamical schemes\cite{wstru00,ashab05,smani06}).
In this case, the analysis above mostly carries through, with the exception that,
since $\tavg{f(\hat H_A) \hat H_{AB}} \ne 0$,
an extra amount of energy is added during coupling, but not removed during
measurement.  This extra energy contributes to the work done on the system
according to Eq.~\ref{e:QW}.  However, the connection to heat found here is
very different because, as the next subsection shows, the definition of heat in
Eq.~\ref{e:QW} requires that the reservoir be near equilibrium.
The comparison presented here is conceptually simpler because energy
stored in the system cannot be instantaneously altered by an external source.

  For a specific example, consider the energy exchange process taking
place between a nuclear spin and its environment in an NMR
spin-relaxation experiment.\cite{ldios11}
In order to represent stored energy, the Hamiltonian of the atom can be defined
with respect to some static field, $\hat H_A = \frac{\hbar \omega_0}{2} \sigma_z$.
Rather than varying the field strength directly,
changing the atomic state from its initial equilibrium
can be brought about with an interaction Hamiltonian, such as the JCM studied here.
The work can be added over each time interval to give,
\begin{equation}
\int^t_0 dt'\; W(t') = \frac{\hbar\omega_0}{2} \Tr{\sigma_z (\rho_A(t) - \rho_A(0))} - \int_0^t dt'\; Q(t')
.
\end{equation}
The heat release can be analyzed using either of the methods in the next section (Sec.~\ref{s:QB}).
Assuming the minimum heat release leads to $\int_0^t dt' \beta(t')Q(t') = S_A(t)-S_A(0)$,
in agreement with the rules of equilibrium thermostatics.
Alternately, in the limit where the $B$ system always begins at thermal equilibrium
and moves infinitesimally slowly between each measurement interval,
Eq.~\ref{e:QW} is recovered, giving $W(t) = 0$.

\subsection{ Comparison to Common Approximations for the Heat Evolution}\label{s:QB}

  The heat generated in the process of Figs.~\ref{f:ref} and~\ref{f:therm}
comes directly from the entropy
change of the measurement system, $B$.
Most analyses ignore the measurement system, making this result
difficult to compare with others in the literature.
Here I present two simple methods for calculating $\Delta S_B$
from quantities available in other methods.

  First, assuming the time-dependence of $\rho_A(t)$ is known,
a lower bound on the heat emitted can be derived
from the state function, $S_A(t) = -\Tr{\rho_A\log\rho_A}$.
Because over each time interval, $\Delta S_A + \Delta S_B \ge 0$,
the total heat added obeys the inequality,
\begin{equation}
\Delta Q(t) = -\Delta S_B/\beta \le \Delta S_A/\beta
.
\end{equation}
Assuming the minimum required heat release leads to a prediction
of the quasistatic heat evolution,
\begin{equation}
\int_0^t \frac{dQ(t')}{dt'} dt' \le \int^t_0 dS_A(t')/\beta(t') \; dt'
.
\end{equation}
This is exactly the result of equilibrium quantum thermodynamics,
valid for arbitrary processes, $\rho_A(t)$.

  Second, if the $B$ system always begins in thermal
equilibrium, $\rho_B(0) = \rho_B^{(\beta)}$,
and the change in occupation probability for each energy level
($\Delta \operatorname{diag}(\rho_B)$) over a
measurement interval is small, then we can directly
use the expansion,\cite{hspoh78}
\begin{equation}
\delta S_B = -\sum_j \delta p_j \log p_j \label{e:dSB}
\end{equation}
This is helpful because in Fig.~\ref{f:therm}, the entropy of the $B$ system
is always calculated in the energy basis of $B$.
Substituting the canonical equilibrium distribution,
\begin{equation}
\delta Q = - \sum_j \delta p_j E_j = -\delta H_B \label{e:dQB}
.
\end{equation}
Equations~\ref{e:dSB} and~\ref{e:dQB} apply whenever
$\rho_B(0)$ is a canonical distribution and
the change in $\rho_B$ is small over an interval.

  In the van Hove limit (Sec.~\ref{s:weak}), energy is conserved between
the $A$ and $B$ systems.
Because of energy conservation, the heat evolution of Eq.~\ref{e:dQB}
is exactly the well-known result of Eq.~\ref{e:QW} in this case.

\section{ Thermodynamic Consistency}\label{s:clausius}

  For the definitions of work and heat given above to be correct, they must meet two requirements.
In order to satisfy the first law, the total energy gain at each step must equal the heat plus work
from the environment.  This is true by construction because the total energy change over each cycle is
just $\tavg{\Delta\hat H_A}$.  Next, in satisfaction of the second law,
the present section will show that there can only be a net heat release over any cyclic process.
Since $Q$ has been defined as heat input to the system, this means
\begin{equation}
\oint Q \le 0.
\end{equation}

  There is a fundamental open question as to whether the energy change caused by
the measurement process should be classified as heat or work.  Counting it as heat
asserts that it is spread throughout the environment in an unrecoverable way.
Conversely, counting it as work asserts that measurement can only be brought about
by choosing to apply a stored force over a distance.  In the cycle of Fig.~\ref{f:therm}, it is
classified as work, because this is the only assignment consistent with thermodynamics.

  Counting $\tavg{\gamma \hat H_{AB}}$ as heat leads to a
systematic violation of the second law, as I now show.
Integrating the quantity,
\begin{equation}
R = \avg{\Delta H_A} + \avg{\Delta H_B} - \Delta S_B/\beta
,
\end{equation}
over an entire cyclic process cancels $\tavg{\Delta H_A}$, leaving
\begin{equation}
\oint R = \oint \avg{\Delta H_B} - \Delta S_B/\beta
.\label{e:Rcontrib}
\end{equation}
If the $B$ sub-system starts each interval in thermal equilibrium (Eq.~\ref{e:eqB}),
this is the free energy difference used in Eq.~\ref{e:Srel}.
The Klein inequality then proves the {\em positivity} of each contribution to Eq.~\ref{e:Rcontrib}.
Therefore, over a cyclic process, $\oint R \ge 0$.

  A thermodynamically sound definition is found when counting as part of $Q$ only
the entropy change of the reservoir.  Heat comes into this model because the environment is
responsible for transforming $\rho_B(t)$ back into $\rho_B(0)$.
Using a hypothetical quasistatic, isothermal process to achieve this will require adding a
heat, $Q = (S_B(0) - S_B(t))/\beta = -\Delta S_B$.

  I now show that $\oint \Delta S_B \ge 0$ by considering entropy changes for
the $A$-$B$ system jointly.
At the starting point, the two systems are decorrelated,\cite{tsaga13}
\begin{equation}
S[\rho_A(0)\otimes\rho_B(0)] = S_A(0) + S_B(0)
.
\end{equation}
The time-evolution of this state is unitary, so $\rho_{AB}(t)$ has the same value for the entropy.
However, projection always increases the entropy,\cite{ejayn57a,tsaga13} so
\begin{align}
S[\rho_A(t) \otimes \rho_B(t)] &\ge S[\rho_{AB}(t)]
.
\intertext{The $A$ and $B$ systems in the final state are also decorrelated, proving the statement,}
\Delta S_A + \Delta S_B &\ge 0.
\end{align}
This is quite general, and applies to any measurement time, starting state, and Hamiltonian, $\hat H_{AB}$.
Again, for a cyclic process $A$
must return to its starting point, so $\oint \Delta S_A = 0$,
and $\oint Q \le 0$.

  It should be stressed that the results of this section hold regardless of the
lengths of the measurement intervals, $\{t^{(k+1)} - t^{(k)}\}$.  The choice of
Poisson-distributed measurement times is not justified in every case.  This
is especially true for the physical micromaser, where the measurement times should instead
be Gaussian, based on the cavity transit time for each atom.
Instead, choosing measurement times from a Poisson
distribution mimics the situation where a measurement is brought about from an
ideal, random collision-type process.

\section{ Results}
\subsection{ Analysis of the Micromaser}\label{s:jcm}

  Exact numerical results are known for the micromaser in the rotating wave approximation
-- a single-qbit system $(B)$ in state $e$ or $g$ coupled to a single mode of an
optical cavity ($A$) in a Fock state, $n = 0, 1, \ldots$.\cite{sharo06,hwalt06,sharo13}
The Hamiltonian is known as the Jaynes-Cummings Model (JCM),
\begin{align}
\hat H_A &= \hbar \omega^A (\hat n_A + \frac{1}{2}) \label{e:JCM} \\
\hat H_B &= \frac{\hbar \omega^B}{2} (\pure{e} - \pure{g}) \\
\gamma \hat H_{AB} &= \gamma( a_A^\dagger a_B + a_A a_B^\dagger ) \label{e:Hab}
\end{align}
The rotating wave approximation neglects a term,
\begin{equation}
\gamma \hat H_{AB}' = \gamma ( a_A^\dagger a_B^\dagger + a_A a_B ) \label{e:Hab2}
\end{equation}
in the Hamiltonian causing simultaneous
excitation of the qbit and cavity.  It is usually justified when 
the two frequencies, $\omega^A$ and $\omega^B$, are near resonance.\cite{ejayn58}
\footnote{Note that the atom-field interaction should also contain a diamagnetic term that
is ignored here but may sometimes be grouped with an effective change in $\omega^A$.\cite{mcris93}}
After a time, $t$, the initial state will be in superposition
with a state where the photon has been emitted.

  The ideal 1-photon micromaser can be solved analytically because
the total number of excitations is conserved, and
unitary evolution only mixes the states $\ket{e,n-1}$ and $\ket{g,n}$.
Thus the only allowed transitions are between these two states.
Attempting to define the work done by the excited atom on the field
requires measuring the energy of the atom.
This is physically realized in the micromaser when the atom exits
the cavity.  This will project the environment
into a state with known excitation, $n_B = e$ or $g$.

  The work and ending states (and from those the heat) can all be
neatly expressed in terms of $x(t)$, the average number of photons absorbed
by the atom given that a projective measurement on the atom
is performed at time $t$,
\begin{align}
x(t) &\equiv \sum_{n=0}^\infty p_n(\sigma_g |b_n(t)|^2 - \sigma_e |b_{n+1}|^2) \\
\sigma_e(t) &= \sigma_e(0) + x(t) \\
\sigma_g(t) &= \sigma_g(0) - x(t) \\
\avg{\Delta H_A(t)} &= -\hbar \omega_A x(t) \\
\avg{\Delta H_B(t)} &= \hbar \omega_B x(t)
.
\end{align}
The expression for the transition probability, $|b_n(t)|^2$,
is recounted in the appendix.

  The analytical solution gives an exact result for the heat and work
when a measurement is done at time $t$.
Averaging over the distribution of measurement times will then
give the expected heat and work values over an interval.
In the limit of many measurements ($T/t \to \infty$), this expectation gives the rate
of heat and work per average measurement interval.
Note that for the physical micromaser setup, the interaction time is
set by the velocity of the atom and the cavity size -- resulting
in a narrow Gaussian distribution rather than the Poisson process studied here.

  For a Poisson distribution of interaction times, the averages are easily computed to be,
\begin{equation}
\avg{|b_n(t)|^2} = \frac{1}{2}\left(1 - \frac{\lambda^2 + \Delta_c^2}
         {\lambda^2 + \Delta_c^2 + 4 n \gamma^2/\hbar^2}\right)
         .
\end{equation}
Strong and weak-coupling limits of this equation give identical first-order terms,
\begin{equation}
\lim_{\lambda \to \infty} \avg{|b_n(t)|^2} = \lim_{\gamma/\hbar \to 0} \avg{|b_n(t)|^2}
= \frac{2 n\gamma^2/\hbar^2}{\lambda^2 + \Delta_c^2}
.
\end{equation}

  Since measurements happen with rate $\lambda$, the effective rate of atomic
absorptions in these limits is,
\begin{equation}
\lambda\avg{x} = \frac{2\lambda\gamma^2/\hbar^2}{\lambda^2 + \Delta_c^2} \left( \sigma_g \avg{n}  - \sigma_e \avg{n+1} \right) \label{e:dx}
\end{equation}  
this recovers Einstein's simple picture of
photon emission and absorption processes occurring with equal rates,\cite{aeins16}
\begin{align}
dW_\text{abs}/\Delta E_\text{abs} &= \sigma_g B^e_g \avg{n} dt \\
dW_\text{em}/\Delta E_\text{em} &= \sigma_e \left( A^g_e + B^g_e \avg{n} \right) dt
.
\end{align}
All the $A,B$ coefficients are equal to the prefactor of Eq.~\ref{e:dx}
here because $x(t)$ counts only a single cavity mode at frequency $\omega^A$.
In a blackbody, the $A$ coefficient goes as $\omega^2 d\omega$
because more modes contribute.\cite{epowe93}

  The denominator, $\lambda^2 + \Delta_c^2$ is exactly the one that
appears in the traditional expression for a Lorentzian line shape.
Here, however, the measurement rate, $\lambda$ appears rather than
the inverse lifetime of the atomic excited-state.
The line broadens as the measurement rate increases,
and the atom is able to absorb/emit
photons further from its excitation frequency.  Only the
resonant photons will cause equilibration, while others will cause noise.
In the van Hove limit, $\gamma,\lambda \to 0$ and  the contribution of the
resonant photons will dominate.

\begin{figure}
\includegraphics[width=0.45\textwidth]{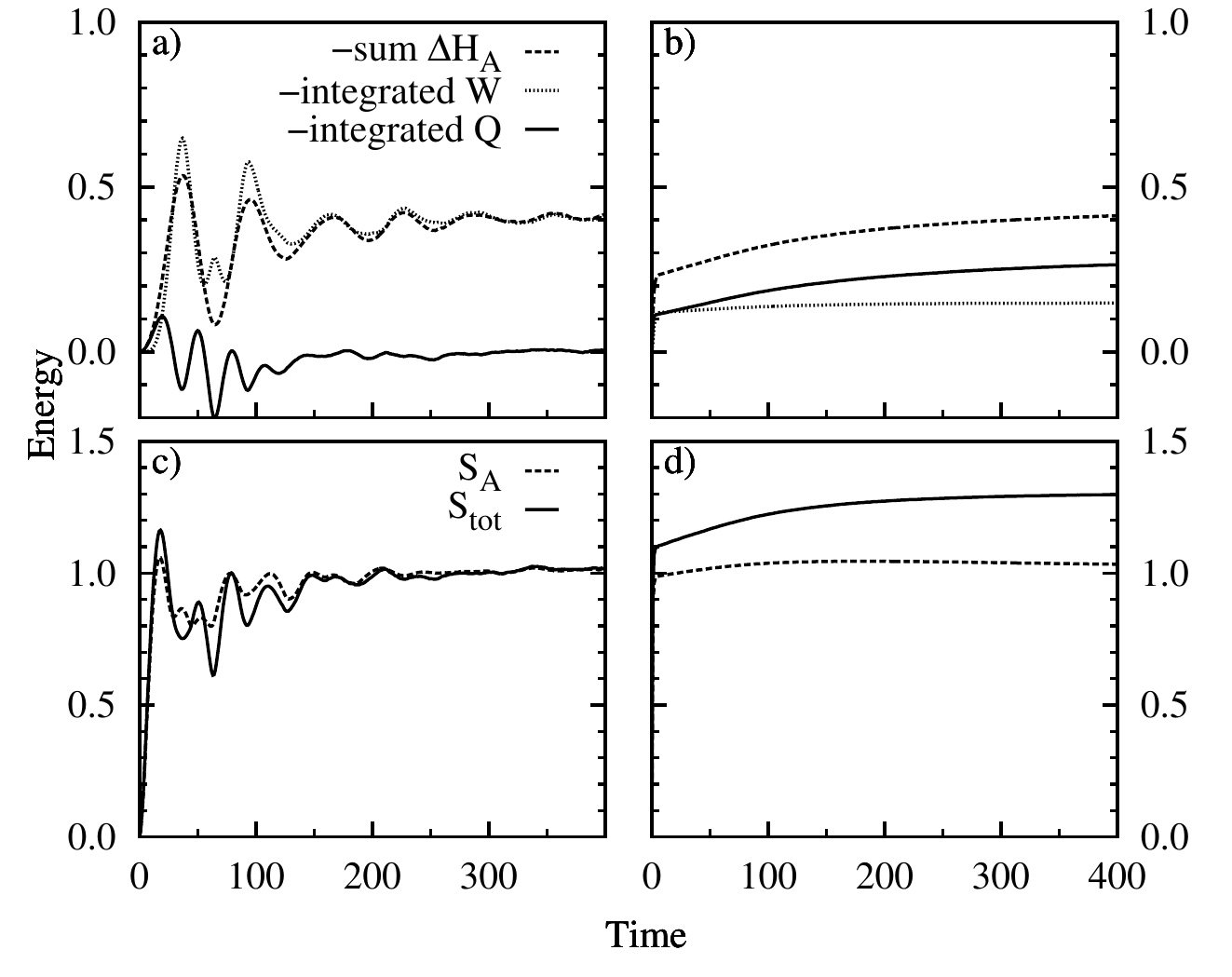}
\caption{Work and heat production during decay of a photon in a cavity ($n_A = 1$) coupled
to a 2-level reservoir (Eqns.~\ref{e:JCM}-\ref{e:Hab2}, $\omega_A = \omega_B = 2\pi$, $\gamma = 0.05$,
$\lambda = 10^{-2}$, $\beta = 1$).
Panels (a) and (b) compare the system energy
loss, $\Delta H_A$, to the work and heat computed from the measured reservoir states
(Eq.~\ref{e:W} and~\ref{e:Q}).  Panels (c) and (d) show the information entropy of
the $A$ system and the combined entropy change, $S_\text{tot}(t) = S_A(t) - \int_0^t Q/T > 0$.
Note that the traditional calculation of heat (Eq.~\ref{e:QW}) gives only
$Q \approx \Delta H_A$, $W \approx 0$.  Panels (a) and (c) show results for the time-evolution
of the density matrix using the exact process, while panels (b) and (d) are computed
using the weak-coupling approximation of Sec.~\ref{s:weak}.
}\label{f:decay}
\end{figure}

  This simple picture should be compared to the full (Rabi) coupling, Eq.~\ref{e:Hab} plus Eq.~\ref{e:Hab2}.
The remaining figures show numerical results for the simulation
of a resonant cavity ($A$) and qubit ($B$) system starting from a cavity in the singly excited energy
state.\cite{jjoha13}
Figure~\ref{f:decay}a compares the average work and heat computed using this cycle
for at the state-point ($\omega^A = \omega^B = 2\pi$, $\gamma/\hbar = 0.05$, $\lambda = 10^{-2}$).
The average was taken over 5000 realizations of process~\ref{f:therm}.
Rabi oscillations can be seen clearly as the photon exchanges with the reservoir (atom).
Initially, this increases the entropy of the incoming atom's energy distribution.
When there is a strong probability of emission, however, the integrated heat release, $-\int_0^t Q(t') dt'$,
shows system actually decreases the entropy of the reservoir.  This happens because
the the reservoir atom is left in a consistent, high-energy, low-entropy state.
In this way, the reservoir can extract useful work from the cavity.
Panel (b) shows that no laws of thermodynamics are broken, since the system starts in a pure state,
but ends in an equilibrium state.  The information entropy of the system itself increases appreciably
during the first Rabi cycle.
Eventually, the equilibration process ends with the initial excitation
energy being transformed into both heat and work.  Despite
the appearance of Fig.~\ref{f:decay}a (which happens for this specific coupling strength),
the emitted heat is generally non-zero.

  The work and entropy defined by Eq.~\ref{e:QW} differ from the results of this section.
Because the earlier definition is based only on the system itself, without considering the reservoir,
there is no way to use the energy of the interacting atom for useful work.
Eq.~\ref{e:QW} therefore finds zero work, and classifies $\Delta H_A$ entirely
as heat lost to the environment.  Panels (b) and (d) of Fig.~\ref{f:decay} show
results from considering the system and reservoir jointly in the weak-coupling
limit as will be discussed in Sec.~\ref{s:weak}.

\subsection{ Weak Coupling Limit}\label{s:weak}

  The classical van Hove limit was investigated in detail by Spohn and Lebowitz,\cite{hspoh78}
who showed generally that thermal equilibrium is reached by $\rho_A$ in this limit
irrespective of the type of coupling interaction, $\hat H_{AB}$.
First, the interaction strength, $\gamma$, must tend to zero so that
only the leading-order term in the interaction remains.  This makes
the dynamics of $\rho_A(t) = \Tri{B}{\rho_{AB}(t)}$ expressible in terms
of 2-point time-correlation functions for the reservoir.
  Next, the long-time limit (here $\lambda \to 0$) is taken
by finding the average density of $\rho_A$ upon measurement.
This enforces energy conservation because time evolution causes
off-diagonal matrix elements to oscillate and
average to zero over long enough timescales.

  Finally, the Gibbs ensemble is found to be stationary by combining
energy conservation with the detailed balance condition
obeyed by the reservoir,
\begin{align}
\Tri{B}{e^{-\beta \hat H_B} \hat A(0) \hat B(t)} &= \Tr{e^{-\beta \hat H_B} \hat B(t-i\beta) \hat A(0)}, \\
\intertext{which enforces for the $A$ system,}
e^{-\beta E^A_n}  B^m_n &= e^{-\beta E^A_m} B^n_m
.
\end{align}
The time-dependence of the operators in this equation is defined
by the Heisenberg picture, below.

  Because the present analysis requires expressions for the time-dependence
of both $\rho_A$ and $\rho_B$, this section re-derives the weak-coupling
limit without taking the partial trace.
The time-dependence of $\rho$ can be found from
second-order perturbation theory,
\begin{align}
\theta_{AB}(t) &= \rho_{AB}(0)
  -\frac{i\gamma}{\hbar}\int_0^t dx\; [\hat H_{AB}(x), \rho_{AB}(0)]
  + O(\tfrac{\gamma^3}{\hbar^3}) \notag \\
&- \frac{\gamma^2}{\hbar^2}\int_0^t ds \int_0^s dx \;
[\hat H_{AB}(s), [\hat H_{AB}(x), \rho_{AB}(0)]]
,\label{e:weak}
\end{align}
where
$\rho_{AB}(0) =  \rho_A\otimes \rho_B(0)$.
This equation uses the following notation for the density matrix
and time-dependence in the interaction representation,
\begin{align}
\theta_{AB}(t) &= U_0^{-t} \rho_{AB}(t) U_0^t \\
\hat H_{AB}(t) &= U_0^{-t} \hat H_{AB} U_0^{t} \\
\intertext{with time-evolution operator,}
U_0 &= e^{-i(\hat H_A + \hat H_B)/\hbar}.
\end{align}

  The time-evolution can be written more explicitly by decomposing
$\hat H_{AB}$ into transitions between joint system/reservoir states ($m$ to $n$)
with energy difference $\omega_n - \omega_m$,
\begin{align}
\hat H_{AB}(t) &= \sum_{\omega} \hat V_\omega e^{i\omega t} \\
\intertext{where}
\hat V_\omega &\equiv \sum_{n,m \;:\; \omega_n - \omega_m = \omega} \pure{n} \hat H_{AB} \pure{m}
.
\end{align}

  It is easy to average each term in Eq.~\ref{e:weak} over Poisson-distributed
measurement times to find,
\begin{align}
\avg{\theta(t)} &= \lambda \int_0^\infty dt\; e^{-\lambda t} \theta(t) \\
 &= \rho_{AB}(0) -\frac{i\gamma}{\hbar}[\tilde H_{AB}(\lambda), \rho_{AB}(0)] + \frac{\gamma^2}{\hbar^2} L' [\rho_{AB}(0)],
 \intertext{where,}
 \tilde H_{AB}(\lambda) &= \sum_\omega \frac{1}{\lambda - i\omega} \hat V_\omega \\
L' \rho &= \sum_{\omega,\omega'} s_{\omega,\omega'} \left(
\hat V_{\omega} \rho \hat V_{\omega'}^\dagger - \frac{1}{2}\{\hat V_{\omega'}^\dagger\hat V_{\omega}, \rho\}
\right) \notag \\
&\qquad\qquad\qquad + \frac{ia_{\omega,\omega'}}{2} [\hat V_{\omega'}^\dagger \hat V_{\omega}, \rho] \label{e:wdiss}
 \\
s_{\omega,\omega'} &= \frac{2\lambda - i(\omega - \omega')}{d_{\omega,\omega'}} \\
a_{\omega,\omega'} &= \frac{\omega + \omega'}{d_{\omega,\omega'}} \\
d_{\omega,\omega'} &= (\lambda - i\omega)(\lambda + i\omega') (\lambda - i(\omega-\omega'))
.
\end{align}
Note that the sums run over both positive and negative transition frequencies, $\omega$, and that
these quantities have the symmetries, $\hat V_{\omega}^\dagger = \hat V_{-\omega}$,
$s_{\omega,\omega'}^* = s_{\omega',\omega}$, $d_{\omega,\omega'}^* = d_{\omega',\omega}$,
and $a_{\omega,\omega'}^* = a_{\omega',\omega}$.
The canonical Lindblad form can be obtained by diagonalizing the matrix, $[s_{\omega,\omega'}]$.

  When $\lambda \to 0$, transitions where energy is conserved between the $A$ and $B$ systems
($\omega = 0$) dominate in the sum, resulting
in a net prefactor of $(\gamma/\lambda\hbar)^2$.  The transition rate
is then $\gamma^2/\hbar^2 \lambda$ -- exactly the combination that is kept constant
in the van Hove limit.
In this limit, tracing over $B$ in Eq.~\ref{e:wdiss}
should recover Eq.~III.19 of Ref.~\citenum{hspoh78}.

  By applying the interaction part of Eq.~\ref{e:wdiss} to the time evolution
with rate $\lambda$, the effective master equation in the weak coupling limit
becomes,
\begin{align}
\pd{\rho_{A}}{t} &= -\frac{i}{\hbar}[\hat H_A, \rho_A(t)] + \frac{\gamma^2 \lambda}{\hbar^2}
\Tri{B}{L' [\rho_A(t)\otimes \rho_B(0)]}
.\label{e:lind}
\end{align}
For the JCM, there is just one $\hat V_{\Delta_c} = a_A a_B^\dagger$, which
gives the same answer as the exact result, Eq.~\ref{e:dx}.

\begin{figure}
\includegraphics[width=0.45\textwidth]{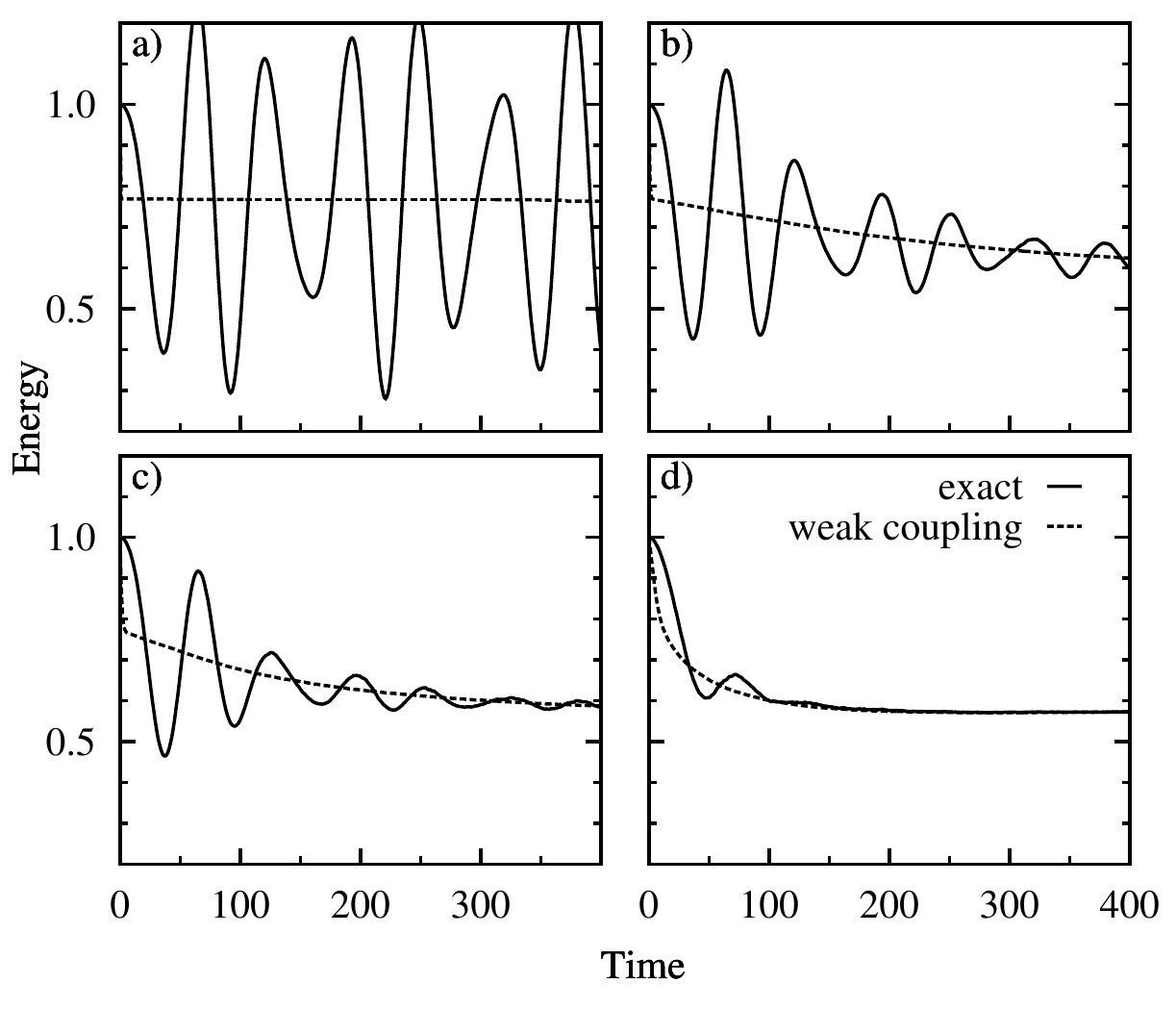}
\caption{Decay of the system simulated in Fig.~\ref{f:decay} from an excited state ($E_A(0) = 1$)
at different values of the measurement rate.  Panels (a)-(d) have rates $\lambda=10^{-4}$,
$5\cdot 10^{-3}$, $10^{-2}$ and $5\cdot 10^{-2}$, respectively.
The exact repeated measurement process is compared with the second-order
perturbation theory of the weak-coupling limit.
The shape of the decay to steady-state behavior is a combination of
fast energy exchange due to Rabi oscillations and the
slower process of memory loss through repeated measurement.}\label{f:cmp}
\end{figure}

  Relaxation process simulated by continuously
applying $L'$ can show qualitative differences from the
process in Sec.~\ref{s:process}.  Without the trace over the environment,
$L'$ just gives the approximation to $\theta(t)$ from
second-order perturbation theory.  This decays faster than
when repeated projection is actually used because the environment
loses its memory after each projection.\cite{ejayn57a}
These two time-scales can be seen in Fig.~\ref{f:cmp}.

  Fig.~\ref{f:cmp} (and Fig.~\ref{f:decay}b,d) compares simulation of $L'$ with the exact
process~\ref{f:ref} when repeated projection is used in the same way for both.
That is, time evolution under the Lindblad equation (\ref{e:lind}) is carried
out in intervals, $t \sim $Poisson($\lambda$).  After each interval,
the purification operator (Eq.~\ref{e:pure}) is applied to the density matrix.
This way, the only difference from the exact process is that the time-propagator
has been approximated by its average.  It is evident that the initial
$\cos^2$ shape and Rabi oscillation structure have been lost.
Instead, the $L'$ propagator shows a fast initial loss followed by simple exponential decay toward
the steady-state.  Nevertheless, the observed decay rate and eventual
steady states match very well between the two methods.
The total evolved heat shows a discrepancy because the fast initial loss in
the $L'$ propagator quickly mixes $\rho_B$.
Numerical simulations of the Lindblad equation were carried out
using QuTiP.\cite{jjoha13}

\subsection{ Fast Coupling Limit}\label{s:strong}

  For the atom-field system, it was shown that the transition rate
approached the same value in both the weak coupling and infinitely
fast measurement case.
To find the general result for the Poisson measurement process as $\lambda \to \infty$,
note that the Taylor series expansion of the
time average turns into an expansion in powers of $\lambda^{-1}$,
\begin{equation}
\lambda \int_0^\infty dt\; e^{-\lambda t} \theta(t) = \sum_{k=0}^\infty \lambda^{-k} \theta^{(k)}(t)
.
\end{equation}

  It is elementary to calculate successive derivatives, $\theta^{(k)}$, by plugging into
\begin{equation}
\pd{\theta(t)}{t} = -\frac{i\gamma}{\hbar} [\hat H_{AB}(t), \theta(t)]
.
\end{equation}
The average measured $\theta$ after a short interaction time on the order of $\lambda^{-1}$
is therefore,
\begin{align}
\avg{\theta} &= \rho_{AB}(0) - \frac{i\gamma}{\lambda \hbar} [ \hat H_{AB}, \rho_{AB}(0) ] \notag \\
 &+ \frac{\gamma}{\lambda^2\hbar^2} \left[[\hat H_A + \hat H_B, \hat H_{AB}], \rho_{AB}(0)\right] \notag \\
 &+ \frac{\gamma^2}{\lambda^2\hbar^2} \left(2 \hat H_{AB} \rho_{AB}(0) \hat H_{AB}
       - \{\hat H_{AB}^2, \rho_{AB}(0)\}  \right) \notag \\
 & + O\left(\frac{\gamma^3}{\lambda^3\hbar^3}\right)
 .\label{e:strong}
\end{align}

  We can immediately see that this limit is valid when the measurement rate is
faster than $\gamma/\hbar$ measurements per second.
The $O(\gamma)$ terms are in the form of a time-propagation
over the average measurement interval, $\lambda^{-1}$.
They have only off-diagonal elements, and do not contribute to
$\tavg{\hat H_A}$ or $\tavg{\hat H_B}$.

  The third term has the familiar Lindblad form, which immediately proves
a number of important consequences.  First, all three terms are trace-free and totally positive.
Next, this term introduces dissipation towards a stationary state for $\rho$.
For a system under infinitely fast repeated measurement,
the $O(\gamma)$ terms do not contribute to Tr$_B$, and
the density matrix evolves according to,
\begin{align}
\dot \rho_A(t) = &-\frac{i}{\hbar}[\hat H_A, \rho_A(t)] \notag \\
&- \frac{\gamma^2}{\lambda \hbar^2}
\Tri{B}{ [\hat H_{AB},[\hat H_{AB}, \rho_{A}\otimes \rho_B(0)]] }
.
\end{align}

  A more explicit representation is possible by defining the sub-matrices,
\begin{equation}
[\hat V^{nm}]_{ij} = [\hat H_{AB}]_{in,jm}.
\end{equation}
These have the symmetry, $\hat V^{nm} = \hat V^{\dagger\, mn}$,
so
\begin{align}
-\Big[ &[\hat H_{AB},[\hat H_{AB}, \rho_{A}\otimes \rho_B(0)]] \Big]_{m,m} \notag \\
&= \sum_n p^B_n
2 \hat V^{mn} \rho_A \hat V^{\dagger\, mn}
- p^B_m \{\hat V^{mn} \hat V^{\dagger\, mn}, \rho_A \}
\end{align}

  For the JCM, this gives,
\begin{equation}
\lambda\avg{x} = \frac{2\gamma^2}{\hbar^2\lambda}(\sigma_g\avg{n} - \sigma_e\avg{n+1})
.
\end{equation}

  The stationary state of this system will usually not
be in the canonical, Boltzmann-Gibbs form.
In fact, the prefactor does not depend on the cavity-field energy mismatch,
$\Delta_c$, so it gives atomic transitions regardless of the wavelength of the light.

  This phenomenon is an explicit manifestation of the energy-time uncertainty
principle.  In the long-time limit of Sec.~\ref{s:weak},
energy-preserving transitions dominated over all possibilities.
In the short-time limit of this section, all the transitions contribute equally, and
the energy difference caused by a transition could be infinitely large.
In-between, energy conservation (and convergence
to the canonical distribution) depends directly on the smallness of the
measurement rate, $\lambda$.

\section{ Minimum Achievable Temperature}\label{s:mint}

  Results from simulating the time-evolution of the open quantum system
using  Eq.~\ref{e:wdiss} reveal that even as the reservoir temperature
approaches zero, the probability of the first excited state does not vanish.
In fact, the results very nearly resemble a Gibbs distribution at
elevated temperatures.  As the reservoir goes to absolute zero,
the effective system temperature levels off to a constant, minimum value.

  This section gives both intuitive and rigorous arguments showing
that this is a general phenomenon originating
from work added during the measurement process.
First, observe that the total Hamiltonian, $\hat H$, is preserved
during coupled time-evolution.  When allowed by the
transitions in $\hat H_{AB}$ (i.e. when $[\hat H, \hat H_{AB}] \ne 0$),
a portion of that total energy will oscillate between
$\hat H_A + \hat H_B$ and $\hat H_{AB}$.
Consider, for example, a dipole-dipole interaction,
$\hat H = \hat x_A^2 + \hat p_A^2 + \hat x_B^2 + \hat p_2^2 + \gamma \hat x_A \hat x_B$.
At equilibrium, the individual systems have $\tavg{\hat x} = 0$,
but the coupled system polarizes so that, $\tavg {\hat H_{AB}} < 0$.

  Intuitively, the joint system can be pictured as relaxing to a thermal
equilibrium at an elevated temperature.
The initial density matrix at each restart, $\rho_A(\beta') \otimes \rho_B(\beta)$,
would then look like an instantaneous fluctuation of
\begin{equation}
\rho_{AB}(\beta') = e^{-\beta \hat H} / Z_{AB}(\beta') \label{e:betap}
\end{equation}
where $\tavg{\hat H_{AB}} = 0$ is too high and $\tavg{\hat H_B}$ is too low.

  At steady state, $\tavg{\hat H_A}$ must be the same at the beginning
and end of every measurement cycle.  This allows the equilibrium argument
above to determine $\beta'$ by self-consistency,
\begin{equation}
\avg{\hat H_B(t) - \hat H_B(\beta)} = -\gamma \avg{\hat H_{AB}(t)}
.
\end{equation}
If equilibrium at $\beta' = 1/k_B T'$ is reached by the average measurement time,
then expanding $\tavg{\hat H_B(\beta') - \hat H_B(\beta)}$ yields,
\begin{equation}
\Delta T \simeq \frac{-\gamma\avg{\hat H_{AB}(t)}}{C_{V,B}}
,
\end{equation}
where $C_{V,B}$ is the heat capacity of the reservoir system.

  It is well-known that quantum mechanical degrees of freedom freeze out
at temperatures that are fractions of their first excitation energy ($\Delta E_1$).
Since the heat capacity when $\beta^{-1} < \Delta E_1$ goes to zero, while
the interaction energy should remain nonzero, this intuitive argument
suggests that the temperature of the system cannot go much below $\Delta E_1/k_B$.

  To be more quantitative, $\tavg{\hat H_{AB}(t)}$ can be estimated in the weak
coupling limit from the second-order perturbation theory of Sec~\ref{s:weak}.
This comparison considers the case $\Delta_c = 0$, since
 the stationary state where $\Delta_c \ne 0$
is known to be non-canonical.
Also, the JCM with rotating wave approximation
is too idealistic, since when $\Delta_c = 0$ no off-resonance
interactions can occur -- so $\hat H_{AB}$ commutes with $\hat H$ and the minimum temperature
argument does not apply.  In other words, in the rotating wave approximation,
the number of absorption events, $x(t)$, always increases the energy of the atom
and decreases the energy of the cavity by the same amount.

  However, if the physical interaction Hamiltonian, $\hat H_{AB} = (a_A + a_A^\dagger)(a_B + a_B^\dagger)$
is used, then the weak coupling theory should also include transitions between $0,g$ and $1,e$.
The average number of simultaneous excitations must be tracked separately, since
it increases both the energy of the atom and cavity.  Using Eq.~\ref{e:wdiss}
with $\omega^A = \omega^B = \omega$, this average is
\begin{equation}
\avg{d(t)} = \frac{2\gamma^2/\hbar^2}{\lambda^2 + (2 \omega)^2} \left(
\sigma_g \avg{n+1} - \sigma_e\avg{n}
\right)
.
\end{equation}

  In the low-temperature limit, only the probabilities of the four lowest-lying states,
labeled $p_{0/1}\sigma_{g/e}$, are relevant.
The general result whenever $\hat H_{AB}$ allows
for both $0,e \leftrightarrow 1,g$ and $0,g \leftrightarrow 1,e$ transitions
with with equal weight and respective energy differences of zero
and $2\hbar\omega$ is,
\begin{equation}
\pd{\tavg{\hat H_A}}{t} = \frac{2 \tfrac{\omega}{\lambda} \gamma^2 / \hbar}{(\tfrac{\lambda}{2\omega})^2 + 1}
\left(
(\tfrac{\lambda}{2\omega})^2 (p_0 - p_1) + \sigma_e p_0 - \sigma_g p_1
\right)
.
\end{equation}
This can be solved for steady-state, $\tavg{\hat H_A} = 0$ to find,
\begin{align}
\frac{p_1}{p_0} &= \frac{(\tfrac{\lambda}{2\omega})^2 +\sigma_e}{(\tfrac{\lambda}{2\omega})^2 + \sigma_g}
. \\
\intertext{In the low-temperature limit,}
\lim_{\sigma_g \to 1}  \frac{p_1}{p_0} &= \frac{(\tfrac{\lambda}{2\omega})^2}{(\tfrac{\lambda}{2\omega})^2 + 1}
.\label{e:minT}
\end{align}

\begin{figure}
\includegraphics[width=0.45\textwidth]{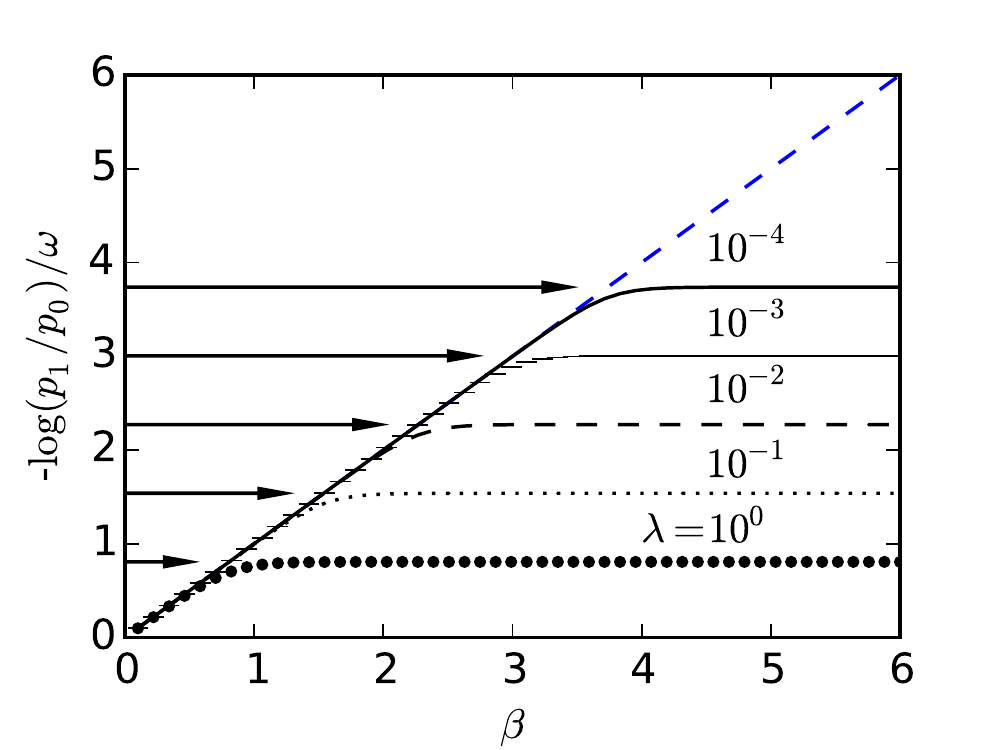}
\caption{Steady-state inverse temperature vs. reservoir $\beta$.
The arrows plot the limiting value of $-\omega^{-1} \log p_1/p_0$ from Eq.~\ref{e:minT}.
Each line represents the steady-states found using a fixed measurement rate, $\lambda$,
as the reservoir temperature varies.
Their y-values were computed from the steady-state
probabilities for simulation in the weak-coupling limit (Eq.~\ref{e:lind}).
}\label{f:steady}
\end{figure}

  This argument brings the energy-time uncertainty principle into sharp focus.
If the measurement rate is on the order of the transition frequency, $\omega$,
then $p_1/p_0$ can be of order 1, making absolute zero unreachable
regardless of the coupling strength, $\gamma$, or the reservoir temperature determining
$\sigma_e/\sigma_g$.  On the other hand, as the relative measurement
rate, $\lambda/\omega$, approaches zero the thermodynamic equilibrium
condition, $\sigma_ep_0 = \sigma_g p_1$, dominates.  In the limit where measurements
are performed very slowly, transitions that do not conserve the energy of the isolated
systems are effectively eliminated.

  Figure~\ref{f:steady} illustrates these conclusions.  For high reservoir temperatures
and low measurement rates, the system's steady-state probabilities follow the canonical
distribution with the same temperature as the reservoir.  When the reservoir
temperature is lowered below a limiting value, the system is unable to respond
-- effectively reaching a minimum temperature determined by Eq.~\ref{e:minT}.
Effects from the minimum temperature can be minimized by lowering the measurement rate.

\section{ Conclusions}

  A measurement process is needed in order to define heat and work a quantum setting.
Continuously measuring the energy of an interacting quantum system
leads either to a random telegraph process or else to the quantum Zeno paradox,
while waiting forever before measuring the energy leads the EPR paradox.
The resolution by intermittent measurement
leads to the conclusion that
quantum systems under measurement do not always reach canonical (Boltzmann-Gibbs) steady-states.
Instead, the steady-state of a quantum system depends both on its coupling to an external
environment and the rate of measurement.

  The presence of a measurement rate in the theory indicates the importance of the
outside observer -- a familiar concept in quantum information.
Most experiments on quantum information have been analyzed in the context of a
Lindblad master equation, whose standard interpretation
relies on associating a measurement rate to every dissipative term.
This work has shown that every dissipative term can be a source/sink for
both heat and work.

  This work has re-derived the master equation in the limit
of weak coupling for arbitrary (Poisson-distributed) measurement rates.
The result agrees with standard line-shape theory, and
shows that measurement rates on the order of the first excitation
energy can cause observable deviations from the
canonical distribution.

  The physical consequences of the measurement rate will become increasingly
important as quantum experiments push for greater control.\cite{bdanj16}
However, they also present a new probe of
the measurement rule and energy-time uncertainty principle for quantum mechanics.
For the micromaser, the rate {\em seems} to be the number of atoms sent through
the cavity per unit time -- since every atom
that leaves the cavity is measured via its interaction with the outside environment.
It is not, however, because even there the atoms can be left isolated and held in a superposition state
indefinitely, leading to entanglement between successive particles.\cite{sharo13}
Most generally, the number of measurements per unit time is determined by
the rate at which information can leak into the environment.
If information leaks quickly, the amount of energy exchanged can be large
and the minimum effective temperature of the system will be raised.
If information leaks slowly, the work done by measurement will be nearly
zero, and the quantum system will more closely approach the canonical distribution.
By the connection to the width of spectroscopic lines, this rate is
closely related to the excited-state lifetime.

  This model presents a novel, experimentally motivated and
thermodynamically consistent treatment of heat
and work exchange in the quantum setting.  By doing so, it also raises
new questions about the thermodynamics of measurement.
First, the explicit connection to free energy and entropy of reservoir states
provides an additional source of potential work that may be extracted from coupling.
Connecting multiple systems together or adding partial
projection using this framework will
provide more realistic conditions for reaching this maximum efficiency.
Second, we have shown special conditions that cause the present definitions to
reduce to well-known expressions in the literature.
Third, although the initial process was defined in terms of wavefunctions,
the average heat and work is defined in terms of the density matrices.
Definitions (Eq.~\ref{e:Q} and~\ref{e:W}) still apply
when the density matrix consists of a single state, but the repeated measurement
projecting to a single wavefunction has a subtly different interpretation.
The difference (not investigated here) is related to Landauer's principle,\cite{swook11,elutz15}
since measuring the exact state from the distribution, $\rho_A \otimes \rho_B$,
carries a separate `recording' cost.

  Stochastic Schr\"{o}dinger equation and power measurement based methods
assume that all energy exchange with the reservoir is as heat.
There, work is supplied by the time-dependence of the Hamiltonian.
As we have shown here, heat is most closely identified with the von Neumann
entropy of the $A$ system.  The energy exchange with the reservoir is
only indirectly connected to the heat exchange through Eq.~\ref{e:dQB}.
The fact that this becomes exact in the van Hove limit explains the
role of the steady-state for $A$ and observations by many
authors that the work of measurement is the source of non-applicability
of fluctuation theorems.\cite{cjarz04,kfuno13,cjarz15,bvenk15,sdeff16}

  When $\Delta H_A + \Delta H_B = 0$, the measurement back-action disappears, and
the fluctuation theorem for $\Delta H_A$ is given by the formalism
of Ref.~\citenum{gmanz15}.
It should also be possible to derive a forward fluctuation theorem
(not restricted to time-reversal) for predicting the force/flux relationships
along the lines of Refs.~\citenum{droge12}.

  There have been many other investigations on thermodynamics of driven, open
quantum systems.  The restriction to time-independent Hamiltonians in this work
differs from most others, which assume a pre-specified, time-dependent
$\hat H_A(t)$.  To make a comparison, either the cycle should be modified as described
in Sec.~\ref{s:issues} or work at each time-step in such
models must be re-defined to count only energy
that is stored in a time-independent Hamiltonian for the central system, $H_A$.

  The process studied here retains a clear connection to the experimental measurement process,
and is flexible enough to compute heat and work for continuous feedback control.
In view of the near-identity between Eq.~\ref{e:minT} and Eq.~10 of Ref.~\citenum{yutsu10},
it is very likely that recent experimental deviations from the fluctuation theorem
are due to the phenomenon of minimum temperature, as well as to
differences between traditional, system-centric, and the present, observational,
definitions of heat and work.

\begin{acknowledgments}
  I thank Brian Space, Sebastian Deffner, and Bart\l{}omiej Gardas for helpful discussions.
This work was supported by the University of South Florida
Research Foundation and NSF MRI CHE-1531590.
\end{acknowledgments}


\appendix
\section{ Explicit Solution for the JCM}

  The solution to the Jaynes-Cummings model under the rotating wave
approximation is well-known.\cite{hwalt06,sharo06,jhoro12}
I summarize it in the notation of this work for completeness.
  For states with $n > 0$ total excitations, the time-evolution operator
decomposes into a $2\times 2$ block-diagonal,\cite{ejayn58}
\begin{align}
\begin{bmatrix} \avg{n-1,e | \psi(t)} \\ \avg{n,g | \psi(t)} \end{bmatrix} &=
e^{-i\omega^A t(n - \tfrac{1}{2})} \\
&\begin{bmatrix}
a_n(t) & b_n(t) \\
b_n(t) & a_n(t)^* \end{bmatrix}
\begin{bmatrix} \avg{n-1,e | \psi(0)} \\ \avg{n,g | \psi(0)} \end{bmatrix}
,\notag
\end{align}
with the definitions,\cite{sharo06}
\begin{align}
\Omega_n &= \frac{2\gamma}{\hbar} \sqrt{n} \\
\Delta_c &= \omega_B - \omega_A \\
\Omega_n'^2 &= \Omega_n^2 + \Delta_c^2 \\
a_n(t) &= \cos(\Omega_n' t/2) - \frac{i \Delta_c}{\Omega_n'}\sin(\Omega_n' t/2) \\
b_n(t) &= -\frac{i \Omega_n}{\Omega_n'} \sin(\Omega_n' t/2)
.
\end{align}

  Starting at $t=0$ from
$\pure{n-1} \otimes \pure{e}$ gives,
\begin{equation}
\rho_{AB}(t) =  \begin{bmatrix} \ket{n-1,e} \\ \ket{n,g} \end{bmatrix}^T
\begin{bmatrix}
|a_n(t)|^2 & -a_n(t) b_n(t) \\
a^*_n(t) b_n(t) & |b_n(t)|^2 \end{bmatrix}
\begin{bmatrix} \bra{n-1,e} \\ \bra{n,g} \end{bmatrix}.
\end{equation}

  Starting at $t=0$ from
$\pure{n} \otimes \pure{g}$ gives,
\begin{equation}
\rho_{AB}(t) =  \begin{bmatrix} \ket{n-1,e} \\ \ket{n,g} \end{bmatrix}^T
\begin{bmatrix}
 |b_n(t)|^2 & a_n(t) b_n(t) \\
-a^*_n(t) b_n(t) & |a_n(t)|^2 \end{bmatrix}
\begin{bmatrix} \bra{n-1,e} \\ \bra{n,g} \end{bmatrix}.
\end{equation}

  Because of the simplicity of this system, measuring the atom also
projects the cavity into a Fock state.  This simplifies the analysis, since
we only need to track the pure probabilities, $p_n$.  Assuming
the incoming atomic states are chosen to be pure $e$ or $g$
at random (with probabilities $\sigma_e$ or $\sigma_g$, resp.),
\begin{align}
p_n(t) = p_n(0)
 &+ |b_{n+1}(t)|^2(\sigma_g p_{n+1} - \sigma_e p_n) \notag \\
  & - |b_n(t)|^2 (\sigma_g p_n - \sigma_e p_{n-1}). \label{e:pt}
\end{align}
Eq.~\ref{e:pt} uses the fact that $b_0 = 0$.

  This master equation has a non-trivial steady-state at
$p_n = p_0 (\frac{\sigma_e}{\sigma_g})^n$.  The existence of this
steady-state, and the fact that the cavity does not have a canonical distribution,
even when the atom does ($\sigma_e/\sigma_g = e^{-\beta \hbar \omega^B}$)
were noted by Jaynes.\cite{ejayn58}  Experimentally, relaxation to the
canonical distribution occurs because of imperfect isolation of the cavity,
which allows thermalization interactions with external resonant photons
and results in a near-canonical (but not perfect) steady state.\cite{hwalt06}
Such interactions could easily be added to the present model, but
for clarity this analysis focuses on
interaction with the single reservoir system, $B$.

\end{document}